\pdfoutput=1
\documentclass[11pt]{article}

\usepackage[T1]{fontenc}
\usepackage[utf8]{inputenc}
\usepackage{amsmath,amssymb}
\usepackage{microtype} 
\usepackage{tgpagella}
\usepackage{helvet}
\usepackage{courier}
\usepackage{sectsty}
\allsectionsfont{\sffamily}

\usepackage{graphicx}
\usepackage{array}
\usepackage{booktabs}
\usepackage{longtable}
\usepackage{caption}
\usepackage{makecell}
\usepackage{lscape}

\usepackage{hyperref}
\hypersetup{colorlinks=true, linkcolor=blue, citecolor=blue, urlcolor=blue}

\providecommand{\citeproctext}{}
\newcommand{\citeproc}[2]{#2}

\newlength{\cslhangindent}
\newlength{\csllabelwidth}
\AtBeginDocument{\renewcommand{\citeproctext}[1]{}}

\newenvironment{posttablenotes}{%
  \par\begingroup\setstretch{1}\scriptsize
  \noindent\emph{Note:}\ %
}{\par\endgroup}

\usepackage[T1]{fontenc}

\usepackage{setspace}
\AtBeginEnvironment{table}{\footnotesize}
\AtBeginEnvironment{longtable}{\footnotesize}

\title{\sffamily\bfseries Trade, Political Distance and the\\ World Trade Organization}
\author{Samuel Hardwick\thanks{The Australian National University. 
Email: \href{mailto:samuel.hardwick@anu.edu.au}{samuel.hardwick@anu.edu.au}}}
\date{21 September 2025}

\begin{document}

\maketitle
\begin{abstract}
\noindent Trade agreements are often understood as shielding commerce from fluctuations in political relations. This paper provides evidence that World Trade Organization membership reduces the penalty of political distance on trade at the extensive margin. Using a structural gravity framework covering 1948--2023 and two measures of political distance --- based on high-frequency events data and UN General Assembly votes, respectively --- GATT/WTO status is consistently associated with a wider range of products traded between politically distant partners. The association is strongest in the early WTO years (1995--2008). Events-based estimates also suggest attenuation at the intensive margin, while UN vote-based estimates do not. Across all specifications, GATT/WTO membership increases aggregate trade volumes. The results indicate that a function of the multilateral trading system has been to foster new trade links across political divides, while raising trade volumes among both close and distant partners.
\end{abstract}

\noindent\textbf{JEL classifications:} F13, F14, F50, C23

\noindent\textbf{Keywords:} international trade, World Trade Organization, political distance, gravity model

\newpage

\section{Introduction}\label{introduction}

The decades following World War II saw rapid global economic expansion
supported by a multilateral trading system that promoted confidence in
openness. The General Agreement on Tariffs and Trade (GATT) and later
the World Trade Organization (WTO) anchored commitments to
nondiscrimination and multilateral rules, aiming to prevent a return to
the protectionist spirals of the interwar period.

Trade has never been insulated from politics: in the Cold War, trade
norms were divided along geopolitical lines, and governments regularly
use trade policy as a tool for strategic ends. At the same time,
commercial ties between states can persist and grow despite political
antagonism. Understanding how the multilateral trading system has
conditioned this politics--trade relationship is important to evaluating
that system's role and, in the shadow of current tensions, to assessing
the capacity of rules and norms to mitigate geoeconomic fragmentation.

This paper examines whether the relationship between political distance
and trade is moderated by multilateral trade agreements. A structural
gravity model is estimated to gauge the association between political
distance and bilateral trade across extensive and intensive margins.
Interaction terms are used to assess whether institutions like the GATT,
WTO, regional trade agreements (RTAs) and various domestic governance
variables reduce this association.

The approach uses two measures of political distance: a high-frequency
events-based index derived from news coverage and a widely used index
based on United Nations General Assembly (UNGA) voting. The former is
processed with a particle filter to address measurement noise, a novel
application of this technique to event data. The analysis spans from
1948 to 2023, incorporates domestic trade in accordance with gravity
model best practice, and allows a close examination of how institutional
factors have conditioned the politics--trade relationship.

The results suggest that GATT/WTO membership reduces the negative
association between political distance and trade at the extensive
margin, most significantly in the early WTO years (1995--2008). This
extensive-margin attenuation of the political distance penalty is robust
across specifications, different measurements and checks such as
excluding China or democratic-only pairs. Using the events-based index,
GATT/WTO accession also weakens the politics--trade link at the
intensive margin.

These findings complement recent work, notably by Jakubik and Ruta
(\citeproc{ref-jakubikTradingFriendsUncertain2023}{2023}), by showing
that multilateral institutions help explain variation in how strongly
trade aligns with geopolitics. In doing so, they contribute new evidence
to long-running debates on the interplay between politics and commerce.

In the 1980s and '90s, several pioneering statistical inquiries
documented a link between cooperative or conflictual political relations
and trade. Pollins
(\citeproc{ref-pollinsConflictCooperationCommerce1989}{1989a},
\citeproc{ref-pollinsDoesTradeStill1989}{1989b}) attributes these
correlations to the positive effects of diplomacy on commerce and to
states' strategic adjustment of trade ties. Gowa and Mansfield
(\citeproc{ref-gowaPowerPoliticsInternational1993}{1993}) hypothesise
that, owing to security externalities, free trade is more likely within
political or military alliances, particularly under bipolar rather than
multipolar orders. Morrow, Siverson, and Tabares
(\citeproc{ref-morrowPoliticalDeterminantsInternational1998}{1998})
find, in contrast, that common interests, proxied by similarity of
alliance portfolios, and joint democracy status are better predictors of
trade than alliance membership per se.

Later work used similar methods to explore further mechanisms. Kastner
(\citeproc{ref-kastnerWhenConflictingPolitical2007}{2007}) shows that
internationalist economic interests moderate the effect of political
tension (measured by UN voting dissimilarity) on commerce with a case
study of mainland China--Taiwan relations. Long
(\citeproc{ref-longBilateralTradeShadow2008}{2008}) examines perceived
conflict risk, finding a negative correlation between conflict
expectations and trade levels.

Armstrong (\citeproc{ref-armstrongPoliticsJapanChina2012}{2012}) applies
a political distance index in a stochastic frontier gravity model and
finds a small but significant effect on import performance. Armstrong's
Japan--China case study describes how China's WTO accession reshaped the
role of politics in trade by giving trading partners confidence in
China's commitment to rules and reform. Q. Chen and Zhou
(\citeproc{ref-chenWhoseTradeFollows2021}{2021}) also find that
friendlier countries tend to import more from each other, and that WTO
membership reduces this link (though possibly not for pairs of
authoritarian states). These studies refine earlier gravity and
simultaneous-equations methods, Q. Chen and Zhou
(\citeproc{ref-chenWhoseTradeFollows2021}{2021}) by adding country-pair
dummies and Armstrong
(\citeproc{ref-armstrongPoliticsJapanChina2012}{2012}) via stochastic
frontier estimation.

Other research focuses on specific diplomatic engagements or
controversies. Nitsch
(\citeproc{ref-nitschStateVisitsInternational2007}{2007}) finds that
official state visits produce a temporary export boost of as much as
8--10\%. Fuchs and Klann
(\citeproc{ref-fuchsPayingVisitDalai2013}{2013}) show that meetings with
the Dalai Lama have preceded sharp, one-year drops in exports to China.
Michaels and Zhi (\citeproc{ref-michaelsFreedomFries2010}{2010})
document a related effect during the 2003 US invasion of Iraq:
US--France tensions reduced US consumption of French (or
French-sounding) products relative to other Eurozone goods within the
same Harmonized System (HS) heading. At the aggregate level, impacts are
less clear. Davis and Meunier
(\citeproc{ref-davisBusinessUsualEconomic2011}{2011}) find no robust
effect of tensions on US--France or China--Japan trade, or on broader US
and Japanese trade and investment flows from 1990 to 2006, hypothesising
that sticky economic ties in a globalised economy restrain the effects
of short-lived political shocks.

While Davis and Meunier
(\citeproc{ref-davisBusinessUsualEconomic2011}{2011}) and Michaels and
Zhi (\citeproc{ref-michaelsFreedomFries2010}{2010}) examine higher
frequencies, most studies use annual data, which has prompted some
concerns about temporal aggregation. Du, Ju, Ramirez, and Yao
(\citeproc{ref-duBilateralTradeShocks2017}{2017}) argue that minor or
moderate political tensions have short-run trade effects that annual
data does not capture adequately. Using monthly trade between China and
major partners (1990--2013) and vector autoregression models, they find
that political shocks can impact trade but typically for less than three
months --- potentially reflecting delays, redirections or temporary
boycotts.

A related strand of research aims to analyse simultaneous causality
between political relations and trade. Polachek
(\citeproc{ref-polachekConflictTrade1980}{1980}) provides evidence from
a simultaneous equations model in support of the liberal peace view
(that trade promotes peace by raising the opportunity cost of conflict).
Reuveny and Kang
(\citeproc{ref-reuvenyInternationalTradePolitical1996}{1996}) find
bidirectional predictive relationships using Granger tests. The same
authors use simultaneous equations for 16 country pairs (1960--1992) and
again find reciprocal links, with signs varying by pair
(\citeproc{ref-reuvenySimultaneousEquationsModelTrade2003}{Reuveny and
Kang, 2003}). Keshk, Pollins, and Reuveny
(\citeproc{ref-keshkTradeStillFollows2004}{2004}) estimate simultaneous
equations and find that conflict depresses trade but trade does not
measurably reduce conflict; Hegre, Oneal, and Russett
(\citeproc{ref-hegreTradeDoesPromote2010}{2010}), using expanded data
and alternative specifications, respond that trade does indeed reduce
conflict. The divergent findings in these studies reflect methodological
challenges, including with exclusion restrictions, functional form and
instruments.

Other studies examine how broader openness affects the link between
trade and conflict. Martin, Mayer, and Thoenig
(\citeproc{ref-martinGeographyConflictsRegional2012}{2012}) theorise
that bilateral openness lowers conflict by raising its opportunity
costs, while multilateral openness (import shares from third countries)
reduces bilateral dependence and thus could \emph{increase} conflict
risk. They provide supporting evidence, especially for geographically
close pairs, from instrumental variables regression. Lee and Pyun
(\citeproc{ref-leeDoesTradeIntegration2016}{2016}) revisit the question
and find that \emph{both} bilateral and multilateral openness reduce
conflict risk, attributing the different result to changes in
specification. F. R. Chen
(\citeproc{ref-chenExtendedDependenceTrade2021}{2021}) argues it is not
multilateral openness per se but extended dependence --- trade with the
allies of a potential target --- that deters conflict, since those
allies can impose economic penalties on the challenger.

This paper also relates to work on how the GATT/WTO has affected trade
margins. Liu (\citeproc{ref-liuGATTWTOPromotes2009}{2009}) decomposes
the effects of accession into intensive and extensive (partner) margins
and finds positive impacts on both. Felbermayr and Kohler
(\citeproc{ref-felbermayrModellingExtensiveMargin2010a}{2010}) conclude
that the WTO has created trade mostly at the intensive margin, while the
GATT increased trade mainly at the partner margin. Dutt
(\citeproc{ref-duttWTONotPasse2020}{2020}) studies the dynamics of trade
after WTO accession and finds growing effects over time, with the
strongest impact on the extensive margin (in this case, share of
products traded).

Kim, Londregan, and Ratkovic
(\citeproc{ref-kimEffectsPoliticalInstitutions2019}{2019}) show that
certain political institutions (such as constraints on the executive)
affect the extensive (product-level) margin but not the intensive margin
of trade, and that GATT/WTO effects vary by sector and predominantly
operate at the extensive margin. They argue that regimes insecure about
political survival may block new trade ties, especially in sectors
involving substantial people-to-people contact, while stable democracies
or consolidated autocracies tend to trade broadly.

The implications that emerge from these studies are that the
GATT/WTO raised world trade overall; in the GATT era, this was
largely via creating new partnerships, with the intensive margin more
salient in the WTO period; and political institutions likely shape
the extensive margin by easing (or not impeding) link formation.

In the present context of rising geopolitical tensions, a widely posed
question is whether trade is reverting to bloc structures aligned with
geopolitical camps. Gopinath, Gourinchas, Presbitero, and Topalova
(\citeproc{ref-gopinathChangingGlobalLinkages2025}{2025}) report early
signs of geoeconomic fragmentation: since 2022, countries in US- or
China-centred blocs (inferred from UNGA voting) have begun trading
relatively less with those in the other bloc. The authors emphasise,
however, that the decoupling is still small in absolute terms and that
`non-aligned' connector economies play a critical bridging role that
limits the global economic impact.

Uncertainty about trade policy shapes these patterns. Jakubik and Ruta
(\citeproc{ref-jakubikTradingFriendsUncertain2023}{2023}) show that
during periods of elevated trade policy uncertainty, countries trade
more with geopolitically aligned partners relative to distant ones.
Without the uncertainty interaction, they find no systematic link
between political distance and trade, suggesting that under sufficient
policy certainty the link is weak. These recent papers
(\citeproc{ref-gopinathChangingGlobalLinkages2025}{Gopinath et al.,
2025}; \citeproc{ref-jakubikTradingFriendsUncertain2023}{Jakubik and
Ruta, 2023}) use theoretically well-founded structural gravity methods,
including country-year and pair fixed effects and, for Jakubik and Ruta
(\citeproc{ref-jakubikTradingFriendsUncertain2023}{2023}), domestic
trade flows.

Motivated by these debates and methodological improvements, this paper
employs a structural gravity framework to examine how the GATT/WTO
moderates the political distance--trade relationship, with domestic
institutional indicators as controls. Results are disaggregated across
extensive and intensive margins, offering insights into the mechanisms
at play. The large panel data available, covering over 160 countries
from 1948 to 2023, help reveal how these dynamics have evolved across
phases of the global trading system.

The remainder of the paper is structured as follows.
Section~\ref{sec-pd} defines the two political distance measures.
Section~\ref{sec-empirical} outlines the empirical strategy and data
employed. Section~\ref{sec-results} unpacks the main results by
intensive and extensive margin with attention to variation over time.
Section~\ref{sec-conclusion} concludes.

\section{Measuring Political Distance}\label{sec-pd}

When included in gravity models, political distance is treated as a
trade resistance. The farther apart countries are politically, the
higher the costs from potentially lower trust, greater uncertainty and
deterred investment. Closer relations, in contrast, could reveal
opportunities for exchange previously unknown or seen as too risky.
Because political distance spans multiple dimensions --- geopolitical
alignment, diplomatic interaction and ideological similarity, for
example --- no single proxy captures it fully. This study uses two
complementary measures: an annual index based on UNGA votes to reflect
longer-term alignment and a high-frequency index from coded events data
to capture short-run shocks and diplomatic engagements.

\subsection{UNGA Ideal Point Distance}\label{unga-ideal-point-distance}

\begin{figure}

\centering

\includegraphics[width=0.9\textwidth]{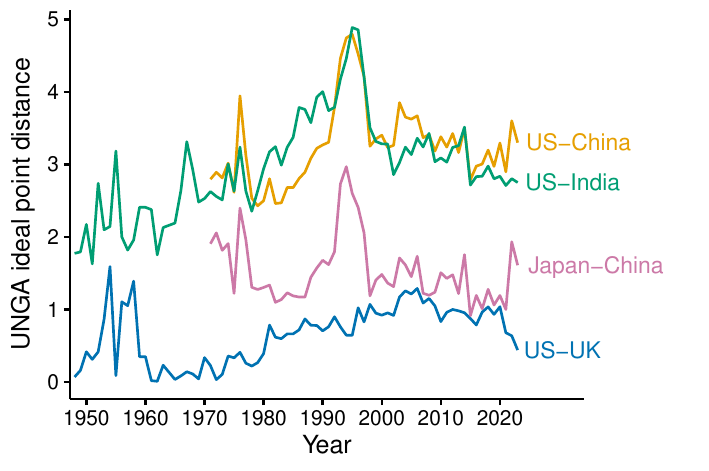}

\caption{\label{fig-ipd}UNGA ideal point distance, selected countries,
1948--2023}

\end{figure}

UNGA roll-call data have been a mainstay for studying foreign policy
alignment. The widely cited ideal point estimates by Bailey, Strezhnev,
and Voeten (\citeproc{ref-baileyEstimatingDynamicState2017}{2017}) are
used here, retrieved from Erik Voeten's Dataverse
(\citeproc{ref-voetenUnitedNationsGeneral2013}{Voeten, Strezhnev, and
Bailey, 2013}). Countries' latent foreign policy positions are estimated
from their voting records with an item-response model fit using Bayesian
methods, where each state's vote is a function of this latent position
and a vote-specific parameter. Repeated votes with identical content
serve as `bridges', with their parameters held fixed to improve
comparability over time. The political distance index used here is the
difference between these two latent positions (the ideal point distance,
shown for selected pairs in Figure~\ref{fig-ipd}).

The series covers all UN members from 1948 onwards, making it well
suited to panel analysis. Because it is based on votes on UNGA agenda
items, it reflects country-specific positions on global issues rather
than bilateral interactions, and since it is based on annual sessions,
it does not capture diplomatic shocks on very short time scales.

\subsection{GDELT Event-Based Index}\label{gdelt-event-based-index}

To complement the ideal point distances, an index is constructed using
the Global Database of Events, Language and Tone (GDELT) 2.0, which
codes news reports using the Conflict and Mediation Event Observations
(CAMEO) taxonomy.\footnote{Coded events data have long been used in
  political science to measure aspects of political relations. Examples
  include Polachek (\citeproc{ref-polachekConflictTrade1980}{1980}),
  Pollins (\citeproc{ref-pollinsConflictCooperationCommerce1989}{1989a},
  \citeproc{ref-pollinsDoesTradeStill1989}{1989b}), Reuveny and Kang
  (\citeproc{ref-reuvenyInternationalTradePolitical1996}{1996},
  \citeproc{ref-reuvenySimultaneousEquationsModelTrade2003}{2003}),
  Davis and Meunier
  (\citeproc{ref-davisBusinessUsualEconomic2011}{2011}), Armstrong
  (\citeproc{ref-armstrongPoliticsJapanChina2012}{2012}), Du et al.
  (\citeproc{ref-duBilateralTradeShocks2017}{2017}) and Q. Chen and Zhou
  (\citeproc{ref-chenWhoseTradeFollows2021}{2021}).} Supported by Google
Jigsaw, GDELT 2.0 draws on print, broadcast and internet news media back
to 1979 (\citeproc{ref-thegdeltprojectGDELT20Event}{The GDELT Project,
n.d.}). Each coded event is assigned a Goldstein score on a
conflict--cooperation scale ranging from --10, representing a military
attack, to +10
(\citeproc{ref-goldsteinConflictCooperationScaleWEIS1992}{Goldstein,
1992}).\footnote{No single event actually scores +10, which would be
  considered commensurate to military conflict. The highest is
  `extending military assistance' at +8.3.}

\begin{figure}

\centering{

\includegraphics[width=\textwidth]{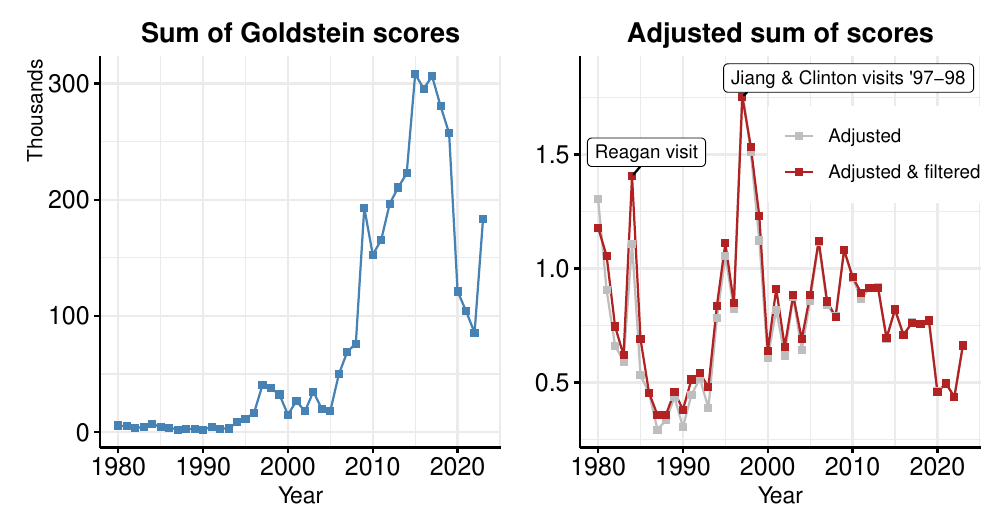}}

\caption{\label{fig-gdelt}Annual GDELT indexes, US--China, 1980--2023}

\end{figure}

Scores are aggregated to the country pair--month level and adjusted for
media coverage by dividing the raw sum of Goldstein scores by the total
number of events reported for either country in that month. Without this
adjustment, large countries in recent years appear more cooperative
simply because they generate more news, producing a secular upward drift
(Figure~\ref{fig-gdelt}). The adjustment normalises by coverage,
yielding a more intuitive series across time and pairs. After the
normalisation, monthly scores must lie within -10 and 8.3; summed
annually, they fall within -120 and 96.

Two practical issues arise with using GDELT-based indexes. First,
temporal aggregation may lead to misleading interpretations: short-lived
shocks affecting trade for a month or two may, once averaged to quarters
or years, give the appearance of an `instantaneous causality' that does
not actually exist (\citeproc{ref-duBilateralTradeShocks2017}{Du et al.,
2017}). While the focus here is on structural correlations rather than
short-run dynamics, robustness checks provided in the appendix
re-estimate the models at a quarterly frequency, including a measure
based on the first month of each quarter (Table~\ref{tbl-robust-gdelt}).
The results are qualitatively unchanged.

Second, coded events may contain errors, such as misidentified actors or
verbs. For sparsely covered pairs, even one or two mistakes could cause
significant distortions. To mitigate this, a latent monthly political
distance series is estimated using a particle filter. The idea is that
countries have a `true' but unobserved political distance, and the
adjusted Goldstein measure acts as a noisy signal of it. Observation
variance is assumed to decline as the number of events rises, reflecting
greater confidence when news coverage is dense. The particle filter was
chosen over other smoothing methods because it does not impose a normal
distribution on the latent series.

The filter works as follows. For the first month in the series, 1000
particles, each representing a potential value of true political
distance, are drawn with replacement from the observed series of summed,
adjusted Goldstein scores. Each particle is then propagated forward from
the previous month by a random walk. The particles are given a weight
proportionate to their likelihood of having generated the observed
adjusted Goldstein measure. They are then re-sampled according to those
weights and averaged to form the final filtered index. A full
description of the process is in the appendix
(Section~\ref{sec-particle}).

Because many country pairs have very sparse event coverage, their
indexes have little meaningful variation. To address this, the baseline
filtered index used in most regressions, requires at least 270 non-zero
months per country pair. This cutoff is simply equal to half the initial
sample window (January 1980--December 2024), and it retains around
one-third (33.6\%) of all possible country pairs. To test sensitivity,
additional results are reported in the appendix with a stricter filtered
index (324 months or 60\%) and with an unfiltered index that includes
all pairs.

The filtering itself mainly affects the early years of the index, where
there are fewer reports of events. For example, as shown in
Figure~\ref{fig-gdelt}, the 1980 US--China score is slightly reduced,
while the 1984 value linked to President Reagan's visit is boosted.
Because higher values correspond to lower political distance, the index
is multiplied by --1 before entering the gravity models to match
expected signs with the UNGA measure.

\section{Empirical Strategy and Data}\label{sec-empirical}

This section outlines the empirical framework used to assess how
institutions attenuate or amplify the relationship between political
distance and trade flows. The analysis primarily uses structural gravity
models
(\citeproc{ref-andersonTheoreticalFoundationGravity1979}{Anderson,
1979}) and the Poisson pseudo-maximum likelihood (PPML)
estimator.\footnote{Estimation is implemented using the \texttt{fixest}
  package for R
  (\citeproc{ref-bergeEfficientEstimationMaximum2018}{Bergé, 2018}).}
PPML is robust to heteroskedasticity and allows the inclusion of zero
trade values (\citeproc{ref-santossilvaLogGravity2006}{Santos Silva and
Tenreyro, 2006}). The empirical strategy is similar to that of Jakubik
and Ruta (\citeproc{ref-jakubikTradingFriendsUncertain2023}{2023}), who
also use interaction terms with political distance in a structural
gravity setting.

Bilateral trade flows \(X_{ijt}\) are modelled as a function of political
distance and institutional variables:
\begin{align*}
    X_{ijt}= \exp \biggl( \bigl[ \beta_0 (PD_{ijt})
      + \boldsymbol{\beta}_{z}^{\top} (PD_{ijt}\times \mathbf{z}_{ijt}) \bigr]\,
      \mathbf{1}_{(i \neq j)} \\
      + \delta_{it}+\delta_{jt}+\delta_{ij}+Border_{ijt} \biggr)
      \times \varepsilon_{ijt},
\end{align*}
where \(PD_{ijt}\) is political distance, \(\mathbf{z}_{ijt}\) is a
vector of institutional variables and \(\mathbf{1}_{(i \neq j)}\) ensures
the political distance term is excluded for domestic trade.

The fixed-effects structure follows current standards in the current
gravity literature. Exporter--year (\(\delta_{it}\)) and importer--year
(\(\delta_{jt}\)) effects absorb country-specific, time-varying shocks
(such as unilateral liberalisation) and capture multilateral
resistances. Country-pair effects (\(\delta_{ij}\)) absorb relatively
time-invariant bilateral characteristics such as geography or
language.\footnote{Larch and Yotov
  (\citeproc{ref-larchEstimatingEffectsTrade2024}{2024}) note that
  country-pair dummies outperform observables such as distance and
  colonial ties in predicting bilateral trade costs.} A border--year
dummy (\(Border_{ijt}\)) takes the value 1 if the flow is international
and the year is \(t\), which controls for globalisation trends that
affect international and domestic trade differently
(\citeproc{ref-bergstrandEconomicIntegrationAgreements2015}{Bergstrand,
Larch, and Yotov, 2015};
\citeproc{ref-yotovAdvancedGuideTrade2016}{Yotov, Piermartini, Monteiro,
and Larch, 2016}).\footnote{One dummy is dropped to avoid perfect
  collinearity.}

Consecutive-year panel data is used, rather than averages or intervals,
to maximise the information in the sample, as recommended by Egger,
Larch, and Yotov
(\citeproc{ref-eggerGravityEstimationsInterval2022}{2022}).

The main dependent variable is bilateral trade, \(X_{ijt}\), measured in
US dollars. Two sources are used. IMF Direction of Trade Statistics
(DOTS) provide long temporal coverage (annual trade from 1948) and are
also available at monthly and quarterly frequencies (from 1960). CEPII's
TradeProd database (\citeproc{ref-mayerCEPIITradeProduction2023}{Mayer,
Santoni, and Vicard, 2023}) covers a shorter period (1966--2020) but has
several advantages: it is a balanced panel, disaggregated into 9
industrial sectors and includes domestic trade, consistent with
structural gravity theory
(\citeproc{ref-yotovRoleDomesticTrade2022}{Yotov, 2022}).

Domestic trade for regressions that use IMF DOTS data is approximated as
GDP (from IMF International Financial Statistics) minus exports. To
address missing observations, zeros were inserted for missing values
where both countries had reported trade with other partners in prior
years.

To examine margins of trade, three related dependent variables are used:
(1) \(S_{ijt}\), the number of TradeProd sectors traded (`sectors'); (2)
\(X_{ijt} / S_{ijt}\), average trade per sector in US dollars (`\$ /
sector'); (3) the number of HS-6 products traded, based on CEPII's BACI
database (`HS-6 products').

In addition to PPML, probability models are estimated for
extensive-margin models where the outcome is bounded. The concern with
using PPML to estimate these models is that positive probability may be
assigned to values above the maximum possible number of sectors. For
HS-6 product counts, this is unlikely to matter. The BACI data set
contains over 5000 products; no country pair reaches the maximum and
very few come close.\footnote{The highest observed value is 4946
  products, for US exports to Canada in 1995.} TradeProd sector counts
may present an issue, however, since the dependent variable is capped at
9 and many pairs do in fact trade in all sectors.

Accordingly, logit models are estimated of the form: \[
  \mathbb{E} \left[ Y_{ijt} = 1 \, | \, \cdot \, \right] = \Lambda \bigr[ \beta_0 (PD_{ijt})
      + \boldsymbol{\beta}_z^{\top} (PD_{ijt}\times \mathbf{z}_{ijt}) + \delta_{it}+\delta_{jt}+\delta_{ij}+Border_{ijt}  \bigr] ,
\] where \(Y_{ijt}\) is either the fraction of sectors traded,
\(S_{ijt}/9\), or a dummy for whether the pair trades at all,
\(\mathbf{1}_{(S_{ijt}>0)}\). These models are estimated using the
split-panel jackknife method with bootstrap standard errors
(\citeproc{ref-hinz2021statedependenceunobservedheterogeneity}{Hinz,
Stammann, and Wanner, 2021}). A drawback is that pairs that always or
never trade in the dataset are dropped due to perfect classification,
which shrinks the sample somewhat. The logit models are therefore used
as a supplement rather than a substitute for the PPML estimates.

The institutional variables include indicators for whether one or both
partners are members of the GATT/WTO, whether the pair are party to the
same regional trade agreement (RTA), and indexes for aspects of domestic
governance. Political distance is interacted with each partner's Polity5
score, a widely used index of regime type that ranges from autocracy to
liberal democracy, and the V-Dem political corruption index. These two
variables were selected because they are only moderately correlated with
each other (-0.47), capture different dimensions of institutional
quality, and offer broad temporal and country coverage.

As robustness checks, two alternative governance indexes are used. These
are the Worldwide Governance Indicators (WGI) Voice and Accountability,
and Rule of Law. Because the WGI series start in 1996, models using them
cover a shorter period. Surprisingly, none of the institutional
covariates used are strongly correlated with GATT/WTO status: the
highest correlation coefficient is with the WGI Voice and Accountability
at 0.45, while the lowest in absolute terms is with the V-Dem political
corruption index at -0.29.

Some specifications omit the \(GATTWTO_1\) indicator (equal to 1 if
exactly one country in a pair is a GATT/WTO member). In the logit
models, estimation drops country pairs that never trade due to perfect
classification, leaving few pairs where neither country is in the
GATT/WTO. In the BACI product-count regressions, which begin in 1995,
almost all pair--year combinations have at least one WTO member. In both
cases, the scarcity of pairs with no WTO membership generates
collinearity when both \(GATTWTO_1\) and \(GATTWTO_2\) are included as
regressors. Accordingly, no-member and one-member pairs are effectively
collapsed into the same category, and only the \(GATTWTO_2\) indicator
is retained in these models. This adjustment ensures that joint
membership and its interaction with political distance --- the main
quantities of interest --- can still be examined while avoiding
collinearity issues.

Some explanatory variables (GDELT political distance, Polity and the
WGI) can take negative values, making them unsuitable for direct
logarithmic transformation. The inverse hyperbolic sine (IHS)
transformation, \(\sinh^{-1}(x) = \ln(x + \sqrt{x^2 + 1})\), is applied
in these cases. This function is defined for all real numbers and
preserves the variables' original order.

Summary statistics for the main variables are reported in
Table~\ref{tbl-summary}. A full list of variables, sources and time
coverage is provided in the appendix (Table~\ref{tbl-data}).

\begingroup
\setlength{\tabcolsep}{3pt}
\renewcommand{\arraystretch}{1.05}

\begin{longtable}{@{}%
  >{\raggedright\arraybackslash\footnotesize}p{0.42\linewidth}%
  >{\raggedleft\arraybackslash\footnotesize}p{0.08\linewidth}%
  >{\raggedleft\arraybackslash\footnotesize}p{0.08\linewidth}%
  >{\raggedleft\arraybackslash\footnotesize}p{0.08\linewidth}%
  >{\raggedleft\arraybackslash\footnotesize}p{0.12\linewidth}%
  >{\raggedleft\arraybackslash\footnotesize}p{0.12\linewidth}@{}}
\caption{Summary statistics}\label{tbl-summary}\\
\toprule
Variable & Mean & SD & Min & Max & N \\
\midrule
\endfirsthead

\toprule
Variable & Mean & SD & Min & Max & N \\
\midrule
\endhead

\bottomrule
\endlastfoot

Trade value (IMF, bn USD) & 1.02 & 75.91 & 0 & 25{,}701.16 & 1{,}674{,}321 \\
Trade value (TradeProd, bn USD) & 0.72 & 45.85 & 0 & 13{,}799.04 & 1{,}211{,}097 \\
Number of TradeProd sectors & 4.33 & 3.97 & 0 & 9 & 1{,}211{,}097 \\
Average trade per sector & 0.08 & 5.09 & 0 & 1{,}533.23 & 1{,}211{,}097 \\
Number of HS-6 products & 216.37 & 587.85 & 0 & 4{,}946 & 1{,}195{,}211 \\
Any-trade dummy & 64.1\% & 0.48 & 0 & 1 & 1{,}211{,}097 \\
\(PD\) (GDELT) & 0.14 & 0.42 & -24.68 & 12.84 & 740{,}921 \\
\(PD\) (UNGA) & -0.03 & 0.90 & -3.13 & 3.22 & 1{,}604{,}185 \\
\(GATTWTO_1\) & 40.1\% & 0.49 & 0 & 1 & 1{,}674{,}311 \\
\(GATTWTO_2\) & 49.8\% & 0.50 & 0 & 1 & 1{,}674{,}311 \\
\(RTA\) & 12.8\% & 0.33 & 0 & 1 & 1{,}674{,}321 \\
\(Polity\) & 1.83 & 7.19 & -10 & 10 & 1{,}224{,}790 \\
\(Corruption\) & 0.49 & 0.30 & 0 & 0.97 & 1{,}521{,}664 \\
Voice and Accountability & -0.05 & 1.00 & -2.31 & 1.80 & 916{,}671 \\
Rule of Law & -0.06 & 0.99 & -2.59 & 2.12 & 920{,}235 \\
\end{longtable}
\endgroup

\section{Results and Discussion}\label{sec-results}

In baseline gravity estimates without institutional variables, political
distance is generally negatively associated with trade
(Table~\ref{tbl-pdonly}). The negative coefficient estimate is
significant for three of the four combinations of political distance
measure and trade data source.

\begin{longtable}{@{}%
  >{\footnotesize}l%
  *{4}{>{\footnotesize}c}%
@{}}
\caption{Coefficient estimates, political distance
only}\label{tbl-pdonly}\tabularnewline
\toprule\noalign{}
& (1) & (2) & (3) & (4) \\
\midrule\noalign{}
\endfirsthead
\toprule\noalign{}
& (1) & (2) & (3) & (4) \\
\midrule\noalign{}
\endhead
\bottomrule\noalign{}
\endlastfoot
\(PD\) & --0.014 & --0.059** & --0.013*** & --0.013*** \\
& (0.044) & (0.029) & (0.004) & (0.004) \\
\(N\) & 726,705 & 556,974 & 1,423,480 & 971,763 \\
Trade data & IMF & TradeProd & IMF & TradeProd \\
\(PD\) data & GDELT & GDELT & UNGA & UNGA \\
Years & 1980--2023 & 1980--2020 & 1948--2023 & 1966--2020 \\
\end{longtable}

\begin{posttablenotes}
    Dependent variable is trade in US dollars. All models include importer--year, exporter--year, directional pair and border--year fixed effects. Pair-clustered standard errors in parentheses. * = \(p\) \textless{} 0.1, ** = \(p\) \textless{} 0.05, *** = \(p\) \textless{} 0.01.
\end{posttablenotes}

\vspace{\baselineskip}

For the GDELT index, the association appears to be strongest at the extensive margin: a one-standard-deviation (SD) increase in political distance reduces the number of HS-6 products traded by about 3.9\% (Table~\ref{tbl-pdmargins}). For the UNGA index, the effect is weaker and appears mainly on the intensive margin, with a one-SD increase corresponding to a 0.8\% fall in average trade per sector. The any-trade dummy is essentially unaffected. The results suggest that while political distance is broadly correlated with lower trade flows, this is not universal across specifications and the margin of adjustment depends on the measure used.

\begingroup
\setlength{\tabcolsep}{3pt}
\renewcommand{\arraystretch}{1.05}

\begin{longtable}{@{}%
  >{\raggedright\arraybackslash\footnotesize}p{0.12\linewidth}%
  *{6}{>{\centering\arraybackslash\footnotesize}p{0.13\linewidth}}@{}}
\caption{Effect sizes of one-SD increase in political distance on trade margins}\label{tbl-pdmargins}\\
\toprule
& \makecell{(1)\\\$ value\\(\%)} 
& \makecell{(2)\\\$ / sector\\(\%)} 
& \makecell{(3)\\Sectors\\(\%)} 
& \makecell{(4)\\HS-6 \\ products\\(\%)} 
& \makecell{(5)\\Sector \\ share\\(pp)} 
& \makecell{(6)\\Trade \\ dummy\\(pp)} \\
\midrule
\endfirsthead
\toprule
& \makecell{(1)\\\$ value\\(\%)} 
& \makecell{(2)\\\$ / sector\\(\%)} 
& \makecell{(3)\\Sectors\\(\%)} 
& \makecell{(4)\\HS-6 products\\(\%)} 
& \makecell{(5)\\Sector share\\(pp)} 
& \makecell{(6)\\Trade dummy\\(pp)} \\
\midrule
\endhead
\bottomrule
\endlastfoot

\multicolumn{7}{@{}l@{}}{\textbf{GDELT distance}}\\[2pt]
\(PD\) 
& \makecell{--2.288**\\(1.063)}
& \makecell{--2.181**\\(1.062)}
& \makecell{--2.700***\\(0.137)}
& \makecell{--3.867***\\(0.221)}
& \makecell{--0.313***\\(0.101)}
& \makecell{--0.574**\\(0.237)} \\
\(N\) 
& 556{,}974 & 556{,}974 & 556{,}974 & 522{,}992 & 413{,}294 & 218{,}429 \\
Years 
& 1980--2020 & 1980--2020 & 1980--2020 & 1995--2023 & 1980--2020 & 1980--2020 \\[4pt]

\multicolumn{7}{@{}l@{}}{\textbf{UNGA distance}}\\[2pt]
\(PD\) 
& \makecell{--0.760***\\(0.262)}
& \makecell{--0.801***\\(0.261)}
& \makecell{--0.281***\\(0.072)}
& \makecell{--0.361***\\(0.076)}
& \makecell{--0.076***\\(0.027)}
& \makecell{0.041\\(0.051)} \\
\(N\) 
& 971{,}763 & 971{,}763 & 971{,}763 & 893{,}671 & 820{,}148 & 573{,}173 \\
Years 
& 1966--2020 & 1966--2020 & 1966--2020 & 1995--2023 & 1966--2020 & 1966--2020 \\
\end{longtable}
\endgroup

\begin{posttablenotes}
    Dependent variables (by column): trade value (US dollars), average trade per TradeProd sector, number of TradeProd sectors, number of BACI 6-digit HS goods, share of TradeProd sectors, and a dummy = 1 if trade \textgreater{} 0. Models (1)--(4) use PPML with pair--clustered standard errors; (5)--(6) use logit with split--panel jackknife and 1000 bootstrap replications. All include importer--year, exporter--year and directional pair fixed effects. All except (4) include border--year dummies. * = \(p\) \textless{} 0.1, ** = \(p\) \textless{} 0.05, *** = \(p\) \textless{} 0.01.
\end{posttablenotes}

\vspace{\baselineskip}

Adding institutional variables changes the picture markedly (Table~\ref{tbl-insts}). The GDELT index has a strong negative association with trade flows, with a one-SD increase associated with reduced trade among non-GATT/WTO pairs by about 21\% and among GATT/WTO members by about 7\%. In contrast, UNGA index coefficients lose significance. There is a small but significant  interaction with the importer's corruption level: more corrupt states see a slightly larger distance penalty on average in these models. Curiously, RTA interactions with UNGA distance are negative, suggesting RTAs may promote trade more strongly if parties are in closer geopolitical alignment, or that RTAs among closer pairs are deeper or broader.

\begingroup
\setlength{\tabcolsep}{4pt}
\renewcommand{\arraystretch}{1.05}

\begin{longtable}{@{}%
  >{\raggedright\arraybackslash\footnotesize}p{0.26\linewidth}%
  *{4}{>{\centering\arraybackslash\footnotesize}p{0.16\linewidth}}@{}}
\caption{Coefficient estimates, institutional interactions included}\label{tbl-insts}\\
\toprule
& (1) & (2) & (3) & (4) \\
\midrule
\endfirsthead

\toprule
& (1) & (2) & (3) & (4) \\
\midrule
\endhead

\bottomrule
\endlastfoot

\(PD\) & --0.621*** & --0.606*** & --0.047 & --0.028 \\
       & (0.139)   & (0.116)   & (0.036) & (0.053) \\
\(RTA\) & 0.067**  & 0.092***  & 0.051*  & 0.069** \\
       & (0.032)   & (0.029)   & (0.029) & (0.027) \\
\(GATTWTO_{1}\) & 0.655*** & 0.477*** & 0.334*** & 0.213* \\
       & (0.114)   & (0.094)   & (0.104)  & (0.113) \\
\(GATTWTO_{2}\) & 0.929*** & 0.736*** & 0.643*** & 0.527*** \\
       & (0.128)   & (0.100)   & (0.128)  & (0.134) \\
\(PD \times RTA\) & 0.074 & 0.057 & --0.028** & --0.024*** \\
       & (0.050)   & (0.041)   & (0.012)  & (0.009) \\
\(PD \times GATTWTO_{1}\) & 0.653*** & 0.540*** & 0.057 & 0.076 \\
       & (0.136)   & (0.129)   & (0.037) & (0.054) \\
\(PD \times GATTWTO_{2}\) & 0.537*** & 0.418*** & 0.030 & 0.035 \\
       & (0.138)   & (0.128)   & (0.036) & (0.053) \\
\(PD \times Corruption_{i}\) & --0.028* & --0.021* & --0.004* & --0.004 \\
       & (0.016)   & (0.011)   & (0.002) & (0.003) \\
\(PD \times Corruption_{j}\) & --0.019 & --0.011 & --0.011*** & --0.006** \\
       & (0.017)   & (0.013)   & (0.002) & (0.003) \\
\(PD \times Polity_{i}\) & --0.006 & -0.003 & --0.003 & --0.006*** \\
       & (0.013)   & (0.012)   & (0.002) & (0.002) \\
\(PD \times Polity_{j}\) & --0.003 & 0.019 & --0.002 & --0.007*** \\
       & (0.014)   & (0.012)   & (0.002) & (0.002) \\
\(N\) & 494{,}638 & 436{,}706 & 912{,}522 & 730{,}074 \\
Trade data & IMF & TradeProd & IMF & TradeProd \\
\(PD\) data & GDELT & GDELT & UNGA & UNGA \\
Years & 1980--2020 & 1980--2020 & 1948--2020 & 1966--2020 \\
\end{longtable}
\endgroup

\begin{posttablenotes}
    Dependent variable is trade in US dollars. All include importer--year, exporter--year, directional pair and border--year fixed
    effects. Pair-clustered standard errors in brackets. * = \(p\)
    \textless{} 0.1, ** = \(p\) \textless{} 0.05, *** = \(p\) \textless{}
    0.01.
\end{posttablenotes}

\vspace{\baselineskip}

\begin{figure}

\centering{

\includegraphics[width=0.95\textwidth]{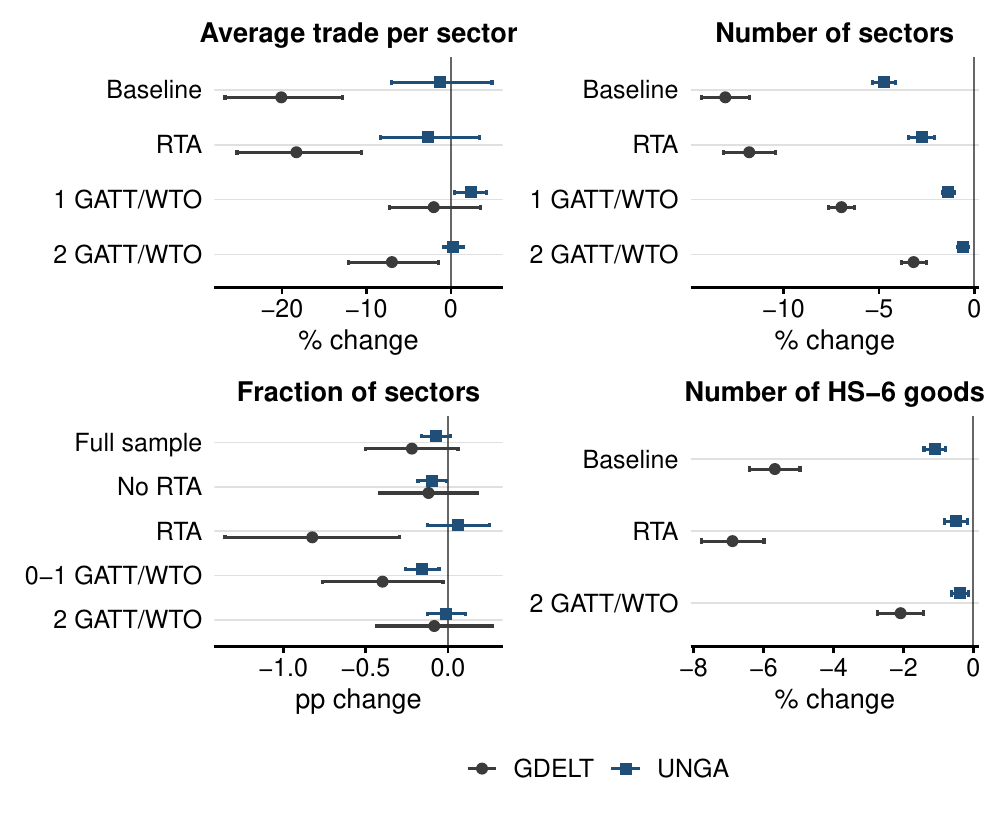}

\caption{\label{fig-margins}Effect sizes of one-SD political distance
shock by margin}

}

\caption*{\noindent \scriptsize \emph{Note:} All estimated with PPML
except fraction of sectors (logit). All models include importer--year,
exporter--year, directional pair and border--year fixed effects. Error
bars show 95\% confidence intervals calculated using the delta method,
based on pair-clustered standard errors (PPML) and bootstrapped standard
errors from 1000 trials (logit).}

\end{figure}

For both GDELT and UNGA indexes, joint WTO membership attenuates the link between political distance and the number of sectors or products traded (Figure~\ref{fig-margins}). For example, in the UNGA models, a one-SD increase in distance is associated with a 1.1\% decline in HS-6 product count for non-WTO pairs but only a 0.4\% decline for WTO members. In the GDELT results, the extensive-margin effect is larger, with sector and product counts both strongly attenuated by WTO membership. Only in the GDELT models is there a substantial intensive-margin interaction. Coefficient tables for each model are provided in the appendix (Table~\ref{tbl-margins-unga} and Table~\ref{tbl-margins-gdelt}).

\begin{figure}

\centering

\includegraphics[width=\textwidth]{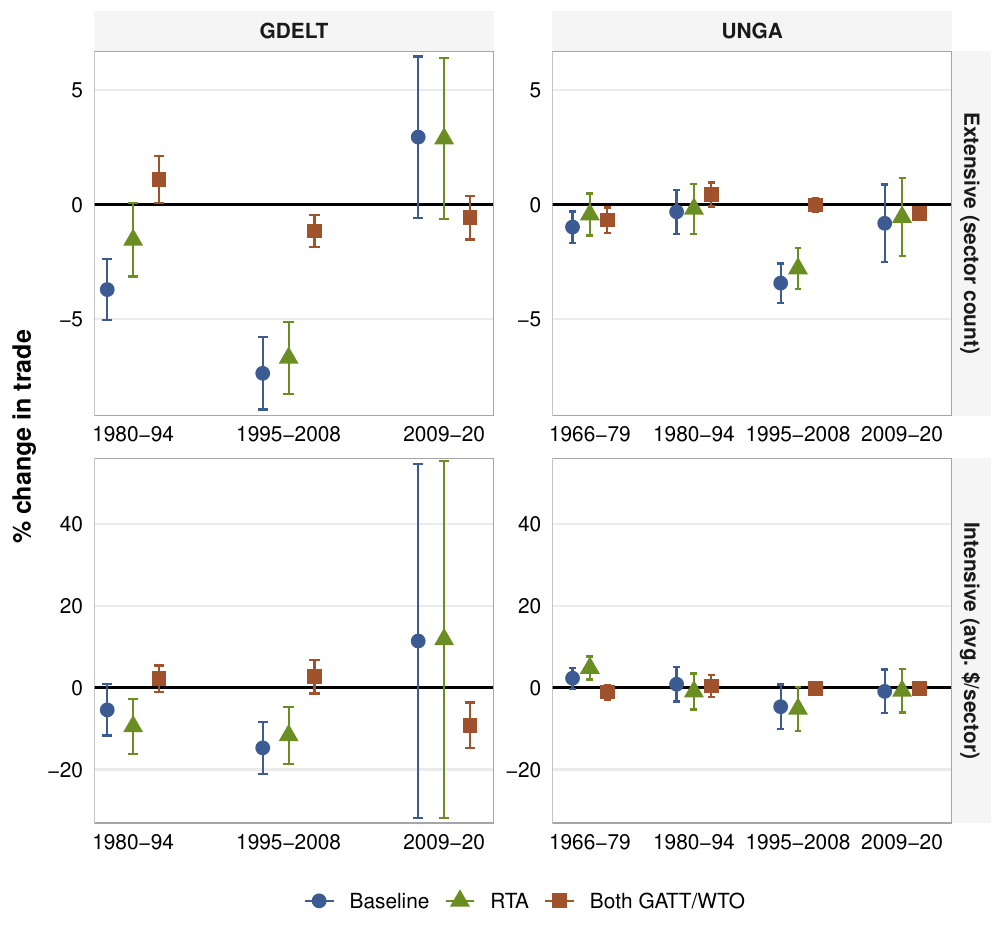}

\caption{\label{fig-hetero}Effect sizes of one-SD political distance shock by margin and subperiod}

\caption*{\noindent \scriptsize \emph{Note:} All models include importer--year, exporter--year, directional pair and border--year fixed effects. Error bars show 95\% confidence intervals calculated using the delta method based on pair-clustered standard errors.}

\end{figure}

Splitting models by period indicates that this attenuation is most
evident in the early WTO years, 1995--2008 (Figure~\ref{fig-hetero}). In
this subperiod, political distance penalties fall for WTO members at the
extensive (product or sector) margin for the UNGA index and at both
margins for the GDELT index. In the post-2008 sample, the GDELT index
shows renewed negative correlations with the intensive margin even among
WTO members, indicative of a world where politics and trade have become
more closely linked. Models without institutional covariates corroborate
this: from 2009 to 2023, GDELT distance is strongly negative, unlike in
earlier periods (Figure~\ref{fig-time-pdonly}).\footnote{Coefficient
  tables for each subperiod are provided in the appendix, disaggregated
  by margin (Table~\ref{tbl-coefs-sectors} and
  Table~\ref{tbl-coefs-avg}) and political distance measure
  (Table~\ref{tbl-gdelt-time} and Table~\ref{tbl-unga-time}).}

\begin{figure}

\centering

\includegraphics[width=\textwidth]{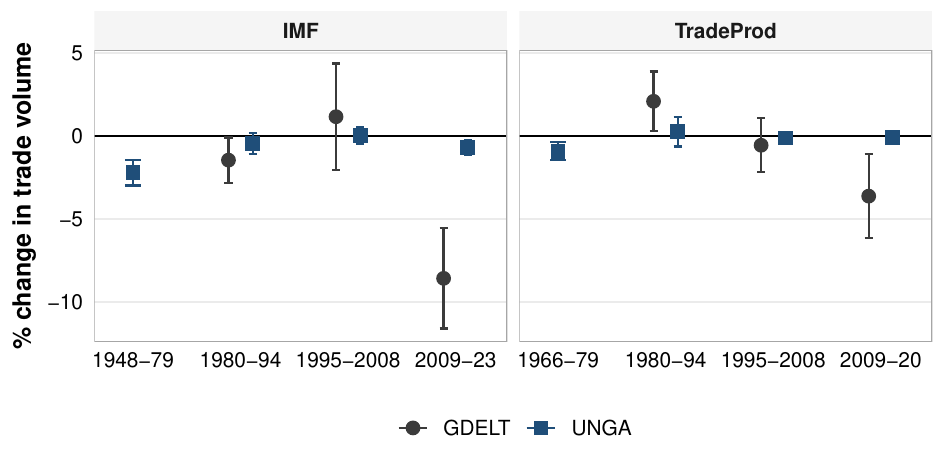}

\caption{\label{fig-time-pdonly}Effect sizes of one-SD political
distance shock by subperiod, political distance only}

\caption*{\noindent \scriptsize \emph{Note:} All models include importer--year, exporter--year, directional pair and border--year fixed effects. Error bars show 95\% confidence intervals based on pair-clustered standard errors.}

\end{figure}

Robustness checks, including dropping China, trimming the top and bottom
1\% of the PD distribution, removing democratic-only pairs and removing
countries at war, do not overturn the core finding of GATT/WTO
membership significantly reducing the distance penalty at the extensive
margin. Quarterly re-estimates with a `first month only' GDELT index
similarly preserve this result. Coefficient tables for these robustness
checks are in the appendix (Table~\ref{tbl-rc-sectors},
Table~\ref{tbl-rc-products} and Table~\ref{tbl-robust-gdelt}).

The analysis points to three broad findings. First, higher political
distance is often but not universally correlated with lower trade flows.
Second, WTO membership appears to have substantially attenuated these
penalties on the extensive margin, consistent with enabling or
sustaining new trade links between politically distant countries. Third, timing matters: the attenuation dynamic was strongest around 1995 to 2008, with
GDELT evidence suggesting that politics and trade have become more
intertwined again since then. RTAs have consistently raised trade
volumes, though their interaction with political distance varies by
margin and by distance measure.

\section{Conclusion}\label{sec-conclusion}

These results offer evidence that WTO membership reduces the negative
association between political distance and trade, though the scope and
strength of this reduction vary across measures and margins. For the
events-based index, this reduced political distance penalty is evident
in aggregate trade flows, while for the UNGA-based index, it appears
only at the extensive (product or sector-level) margin. By subperiod, it
is most significant in the early years of the WTO (1995--2008) when
accessions were frequent and many landmark agreements entered into
force.

Overall, the analysis suggests that a contribution of the multilateral
trading system has been to enable new trade ties across political
divides. While trade volumes did not necessarily grow disproportionately
for politically distant partners compared to close ones, the
extensive-margin result is consistent with the GATT/WTO reducing
politically correlated barriers to entry. That this dynamic appears to
be concentrated in the early WTO years highlights the importance of that
period, marked by the Marrakesh Agreement and other Uruguay Round
outcomes, the strengthened dispute settlement system and the Information
Technology Agreement. Today, the scope for shielding trade from
geopolitical barriers appears narrower and less certain, given the
slowdown in multilateral cooperation, dispute settlement paralysis and
membership saturation, with the WTO counting 166 members as of 2025.

The results here point to several avenues for potential further inquiry.
Firm-level data could help test whether and how multilateral rules
lowered fixed or discovery costs by tracing entry, survival and
upgrading dynamics. Event-study or difference-in-differences designs
could make use of unanticipated changes in political distance, such as
geopolitical realignments in the former Eastern Bloc during the 1990s.
Another prospective line of work, following Gopinath et al.
(\citeproc{ref-gopinathChangingGlobalLinkages2025}{2025}), concerns the
role of connector economies that bridge geopolitical blocs, not only in
sustaining cross-cutting trade ties but in shaping how such links are
initially formed and how institutions mediate them.

\addcontentsline{toc}{section}{References}
\phantomsection

\appendix

\counterwithin{table}{section}
\renewcommand{\thetable}{\Alph{section}\arabic{table}}

\counterwithin{figure}{section}
\renewcommand{\thefigure}{\Alph{section}\arabic{figure}}

\section{Appendix}\label{appendix}

\subsection{Particle Filter}\label{sec-particle}

To address measurement error in the events-based index, a particle
filter (\citeproc{ref-gordonNovelApproachNonlinear1993}{Gordon, Salmond,
and Smith, 1993}) is applied. Compared to other smoothing methods, this
approach has some advantages in the present setting: it does not require
the latent state to be normally distributed, and it can dynamically
adjust the signal-to-noise ratio depending on the density of underlying
events. It handles missing data by propagating particles forward under
the state dynamics when observations are unavailable, and mitigates
distortions from country pairs where coverage is sparse.

The procedure treats `true' political distance as an unobserved latent
state, \(s_{ijt}\), that evolves smoothly over time, while the observed
monthly Goldstein-based series, normalised by coverage, is a noisy
signal, \(y_{ijt}\), of that state. The model is written as: \[
\begin{aligned}
  s_{ijt} &= s_{ij,t-1} + \nu_t, & \nu_t &\sim \mathcal{N}(0, Q_{ij}), \\
  y_{ijt} &= s_{ijt} + \eta_t, & \eta_t &\sim \mathcal{N}(0, R_{ijt}),
\end{aligned}
\] where \(Q_{ij}\) is the process variance, estimated separately for
each country pair \((i,j)\) from an AR(1) fit to the observed
\(y_{ijt}\) series (with a floor of \(10^{-6}\) to ensure positivity).
The observation variance \(R_{ijt}\) is inversely related to the number
of events recorded in a given month, \(n_{ijt}\), so that observations
based on denser media coverage are given more weight:
\(R_{ijt} = \frac{1}{n_{ijt}+1}\).

For each country pair \((i,j)\), the filter proceeds as follows.

First, to initialise the series, \(M = 1000\) particles are generated
for the first month, \(s_{ij,0}\). These particles are sampled from the
empirical distribution of \(y_{ijt}\). Particle weights are initialised
uniformly as \(1 / M\).

Second, each particle \(m\) is propagated forward from \(s_{ij,t-1}\) to
the next month using a random walk: \[
  s_{ijt}^{(m)} = s_{ij,t-1}^{(m)} + \epsilon_{t}^{(m)}, \quad \epsilon_{t}^{(m)} \sim \mathcal{N}(0, Q_{ij}).
\]

Third, the weight of each particle is computed as the Gaussian
likelihood of the observed value \(y_{ijt}\) given the particle and
variance \(R_{ijt}\): \[
  w^{(m)}_{ijt} \propto \phi(y_{ijt} \, | \, s_{ijt}^{(m)}, R_{ijt}),
\] with weights normalised such that they sum to 1.

Fourth, particles are resampled with replacement using the normalised
weights. This discards unlikely particles and duplicates more likely
ones. After resampling, weights are reset to \(1 / M\).

Finally, the filtered political distance index is obtained as the mean
across all resampled particles: \[
  PD_{ijt} = \frac{1}{M} \sum^{M}_{m = 1} s_{ijt}^{(m)} .
\]

The process variance \(Q_{ij}\) is estimated once per country pair as
the residual variance of an AR(1) model fitted to the observed
\(y_{ijt}\) series. If the AR(1) fit fails, a default of 0.001 is used.
Observation variance \(R_{ijt}\) scales down as event counts rise,
improving confidence in well-covered months.

In the baseline index used in regressions, country pairs are retained if
they meet a minimum threshold of 270 months (50\% of the sample window)
of non-zero observations. Another index is used in robustness checks
with a stricter threshold of 324 months (60\%), as well as an unfiltered
index that retains all observations.

\subsection{Additional Tables}\label{sec-tables}

\begingroup
\footnotesize
\setlength{\tabcolsep}{4pt}
\renewcommand{\arraystretch}{1.05}

\begin{longtable}{@{}%
  >{\raggedright\arraybackslash\footnotesize}p{0.20\linewidth}%
  >{\raggedright\arraybackslash\footnotesize}p{0.32\linewidth}%
  >{\raggedright\arraybackslash\footnotesize}p{0.26\linewidth}%
  >{\raggedleft\arraybackslash\footnotesize}p{0.14\linewidth}@{}}
\caption{Variable descriptions, sources and coverage}\label{tbl-data}\\
\toprule
Variable & Description & Source & Years used \\
\midrule
\endfirsthead

\toprule
Variable & Description & Source & Years used \\
\midrule
\endhead

\bottomrule
\endlastfoot

Trade value (IMF) & Bilateral trade flows in current USD. Domestic trade is constructed as GDP minus total exports. & IMF Direction of Trade Statistics (IMF, \citeproc{ref-internationalmonetaryfundimfDirectionTradeStatistics2025}{2025a}); IMF International Financial Statistics (IMF, \citeproc{ref-internationalmonetaryfundimfInternationalFinancialStatistics2025}{2025b}); author's calculations. & 1948--2023 \\
Trade value (TradeProd) & Bilateral and domestic trade flows in current USD. & CEPII TradeProd (\citeproc{ref-mayerCEPIITradeProduction2023}{Mayer et al., 2023}). & 1966--2020 \\
Number of TradeProd sectors traded & Count of up to 9 industrial sectors where bilateral trade is positive. & CEPII TradeProd (\citeproc{ref-mayerCEPIITradeProduction2023}{Mayer et al., 2023}). & 1966--2020 \\
Number of HS-6 products traded & Count of distinct 6-digit HS-1992 products traded. & CEPII BACI (\citeproc{ref-gaulierBACIInternationalTrade2010}{Gaulier and Zignago, 2010}). & 1995--2023 \\
Any-trade dummy & Indicator = 1 if any positive bilateral trade observed. & CEPII TradeProd (\citeproc{ref-mayerCEPIITradeProduction2023}{Mayer et al., 2023}). & 1966--2020 \\
\(PD\) (GDELT) & Political distance index based on GDELT (IHS transformed). & GDELT (\citeproc{ref-thegdeltprojectGDELT20Event}{The GDELT Project, n.d.}); author's calculations. & 1980--2023 \\
\(PD\) (UNGA) & Ideal point distances based on UNGA voting. & Voeten et al. (\citeproc{ref-voetenUnitedNationsGeneral2013}{2013}). & 1948--2023 \\
\(GATTWTO_1\) & Indicator = 1 if exactly one country in the pair is a GATT party or WTO member. & CEPII Gravity (\citeproc{ref-conteCEPIIGravityDatabase2022}{Conte, Cotterlaz, and Mayer, 2022}). & 1948--2023 \\
\(GATTWTO_2\) & Indicator = 1 if both countries in the pair are GATT parties or WTO members. & CEPII Gravity (\citeproc{ref-conteCEPIIGravityDatabase2022}{Conte, Cotterlaz, and Mayer, 2022}). & 1948--2023 \\
\(RTA\) & Indicator = 1 if an RTA is in force between the pair. & WTO (\citeproc{ref-worldtradeorganizationwtoRegionalTradeAgreements2025}{2025}). & 1948--2023 \\
\(Polity\) & Polity5 regime authority spectrum score (IHS transformed). & Center for Systemic Peace (\citeproc{ref-centerforsystemicpeacePolityProject2021}{2021}). & 1948--2020 \\
\(Corruption\) & Political corruption index. & Varieties of Democracy (V-Dem) (\citeproc{ref-coppedgeVDemDatasetV152025}{Coppedge et al., 2025}).\footnote{Retrieved using the \texttt{vdemdata} package for R (\citeproc{ref-maerzVdemdataPackageLoad2025}{Maerz, Edgell, Hellemeier, Illchenko, and Fox, 2025}).} & 1948--2023 \\
Voice and Accountability & Worldwide Governance Indicators (WGI) sub-index (IHS transformed). & WGI (\citeproc{ref-worldwidegovernanceindicatorswgi2024Update2024}{2024}). & 1996--2023 \\
Rule of Law & WGI sub-index (IHS transformed). & WGI (\citeproc{ref-worldwidegovernanceindicatorswgi2024Update2024}{2024}). & 1996--2023 \\
Conflict dyad and location of incompatibility & Countries involved in an armed conflict on opposite sides; location of the incompatibility (for example, over territory) over which the conflict is fought. & UCDP Dyadic Dataset Version 24.1 (\citeproc{ref-daviesOrganizedViolence198920232024}{Davies, Engström, Pettersson, and Öberg, 2024}; \citeproc{ref-harbomDyadicDimensionsArmed2008}{Harbom, Melander, and Wallensteen, 2008}). & 1948--2023 \\
\end{longtable}
\endgroup

\begingroup
\footnotesize
\setlength{\tabcolsep}{4pt}
\renewcommand{\arraystretch}{1.05}

\begin{longtable}{@{}%
  >{\raggedright\arraybackslash}p{0.17\linewidth}%
  *{6}{>{\centering\arraybackslash}p{0.12\linewidth}}@{}}
\caption{Coefficient estimates by margin, UNGA index}\label{tbl-margins-unga}\\
\toprule
& \makecell{(1)\\\$ value} & \makecell{(2)\\\$ / sector} & \makecell{(3)\\Sectors} & \makecell{(4)\\HS-6 \\ products} & \makecell{(5)\\Sector \\ share} & \makecell{(6)\\Trade \\ dummy} \\
\midrule
\endfirsthead

\toprule
& \makecell{(1)\\\$ value} & \makecell{(2)\\\$ / sector} & \makecell{(3)\\Sectors} & \makecell{(4)\\HS-6 \\ products} & \makecell{(5)\\Sector \\ share} & \makecell{(6)\\Trade \\ dummy} \\
\midrule
\endhead

\bottomrule
\endlastfoot

\(PD\) & \makecell{-0.028\\(0.053)} & \makecell{-0.021\\(0.052)} & \makecell{-0.082***\\(0.005)} & \makecell{-0.325***\\(0.005)} & \makecell{-0.022*\\(0.012)} & \makecell{0.007\\(0.023)} \\
\(RTA\) & \makecell{0.069**\\(0.027)} & \makecell{0.071***\\(0.027)} & \makecell{-0.136***\\(0.008)} & \makecell{-0.044***\\(0.008)} & \makecell{0.178***\\(0.051)} & \makecell{0.358***\\(0.132)} \\
\(GATTWTO_{1}\) & \makecell{0.213*\\(0.113)} & \makecell{0.199*\\(0.108)} & \makecell{0.232***\\(0.052)} & --- & --- & --- \\
\(GATTWTO_{2}\) & \makecell{0.527***\\(0.134)} & \makecell{0.505***\\(0.129)} & \makecell{0.462***\\(0.100)} & \makecell{0.206***\\(0.031)} & \makecell{0.375***\\(0.060)} & \makecell{0.370***\\(0.108)} \\
\(PD \times RTA\) & \makecell{-0.024***\\(0.009)} & \makecell{-0.024***\\(0.009)} & \makecell{0.035***\\(0.003)} & \makecell{0.012***\\(0.003)} & \makecell{0.017\\(0.016)} & \makecell{0.037\\(0.038)} \\
\makecell[l]{\(PD \times\) \\ \(GATTWTO_{1}\)} & \makecell{0.076\\(0.054)} & \makecell{0.060\\(0.053)} & \makecell{0.059***\\(0.006)} & --- & --- & --- \\
\makecell[l]{\(PD \times\) \\ \(GATTWTO_{2}\)} & \makecell{0.035\\(0.053)} & \makecell{0.026\\(0.052)} & \makecell{0.071***\\(0.006)} & \makecell{0.027***\\(0.005)} & \makecell{0.030***\\(0.011)} & \makecell{0.028\\(0.021)} \\
\makecell[l]{\(PD \times\) \\ \(Corruption_{i}\)} & \makecell{-0.004\\(0.003)} & \makecell{-0.004\\(0.003)} & \makecell{0.000\\(0.001)} & \makecell{-0.001\\(0.001)} & \makecell{0.012\\(0.008)} & \makecell{0.028\\(0.018)} \\
\makecell[l]{\(PD \times\) \\ \(Corruption_{j}\)} & \makecell{-0.006**\\(0.002)} & \makecell{-0.005**\\(0.003)} & \makecell{0.001\\(0.001)} & \makecell{-0.001\\(0.001)} & \makecell{0.001\\(0.006)} & \makecell{-0.003\\(0.014)} \\
\(PD \times Polity_{i}\) & \makecell{-0.006***\\(0.002)} & \makecell{-0.005**\\(0.002)} & \makecell{-0.001**\\(0.001)} & \makecell{-0.004***\\(0.001)} & \makecell{0.004*\\(0.071)} & \makecell{0.002\\(0.005)} \\
\(PD \times Polity_{j}\) & \makecell{-0.007***\\(0.002)} & \makecell{-0.006***\\(0.002)} & \makecell{-0.001**\\(0.001)} & \makecell{-0.000\\(0.001)} & \makecell{0.000\\(0.002)} & \makecell{0.003\\(0.004)} \\
\(N\) & 730{,}074 & 730{,}074 & 730{,}074 & 504{,}724 & 602{,}191 & 402{,}395 \\
Trade data & Trade-Prod & Trade-Prod & Trade-Prod & BACI & Trade-Prod & Trade-Prod \\
\(PD\) data & UNGA & UNGA & UNGA & UNGA & UNGA & UNGA \\
Years & 1966--2020 & 1966--2020 & 1966--2020 & 1995--2023 & 1966--2020 & 1966--2020 \\
\end{longtable}
\endgroup

\begin{posttablenotes}
    (5) and (6) estimated as logit models using the split-panel jackknife method with bootstrapped standard errors based on 1000 trials. All others estimated using PPML with pair-clustered standard errors. All include importer--year, exporter--year and directional pair fixed effects. All except (4) include border--year dummies. * = \(p\) \textless{} 0.1, ** = \(p\) \textless{} 0.05, *** = \(p\) \textless{} 0.01.
\end{posttablenotes}

\vspace{\baselineskip}

\begingroup
\footnotesize
\setlength{\tabcolsep}{4pt}
\renewcommand{\arraystretch}{1.05}

\begin{longtable}{@{}%
  >{\raggedright\arraybackslash}p{0.17\linewidth}%
  *{6}{>{\centering\arraybackslash}p{0.12\linewidth}}@{}}
\caption{Coefficient estimates by margin, GDELT index}\label{tbl-margins-gdelt}\\
\toprule
& \makecell{(1)\\\$ value} & \makecell{(2)\\\$ / sector} & \makecell{(3)\\Sectors} & \makecell{(4)\\HS-6 \\ products} & \makecell{(5)\\Sector \\ share} & \makecell{(6)\\Trade \\ dummy} \\
\midrule
\endfirsthead

\toprule
& \makecell{(1)\\\$ value} & \makecell{(2)\\\$ / sector} & \makecell{(3)\\Sectors} & \makecell{(4)\\HS-6 \\ products} & \makecell{(5)\\Sector \\ share} & \makecell{(6)\\Trade \\ dummy} \\
\midrule
\endhead

\bottomrule
\endlastfoot

\(PD\) & \makecell{-0.606***\\(0.116)} & \makecell{-0.573***\\(0.114)} & \makecell{-0.358***\\(0.019)} & \makecell{-0.264***\\(0.020)} & \makecell{-0.152**\\(0.062)} & \makecell{-0.351*\\(0.191)} \\
\(RTA\) & \makecell{0.092***\\(0.029)} & \makecell{0.093***\\(0.029)} & \makecell{-0.071***\\(0.007)} & \makecell{-0.055***\\(0.009)} & \makecell{0.141**\\(0.068)} & \makecell{0.130\\(0.199)} \\
\(GATTWTO_{1}\) & \makecell{0.477***\\(0.094)} & \makecell{0.452***\\(0.092)} & \makecell{0.222***\\(0.052)} & --- & --- & --- \\
\(GATTWTO_{2}\) & \makecell{0.736***\\(0.100)} & \makecell{0.700***\\(0.098)} & \makecell{0.428***\\(0.101)} & \makecell{0.233***\\(0.029)} & \makecell{0.384***\\(0.083)} & \makecell{0.583***\\(0.213)} \\
\(PD \times RTA\) & \makecell{0.057\\(0.041)} & \makecell{0.056\\(0.041)} & \makecell{0.037***\\(0.008)} & \makecell{-0.040***\\(0.012)} & \makecell{-0.211***\\(0.070)} & \makecell{-0.137\\(0.194)} \\
\makecell[l]{\(PD \times\) \\ \(GATTWTO_{1}\)} & \makecell{0.540***\\(0.129)} & \makecell{0.521***\\(0.126)} & \makecell{0.173***\\(0.020)} & --- & --- & --- \\
\makecell[l]{\(PD \times\) \\ \(GATTWTO_{2}\)} & \makecell{0.418***\\(0.128)} & \makecell{0.388***\\(0.126)} & \makecell{0.275***\\(0.020)} & \makecell{0.207***\\(0.019)} & \makecell{0.135**\\(0.054)} & \makecell{0.221\\(0.157)} \\
\makecell[l]{\(PD \times\) \\ \(Corruption_{i}\)} & \makecell{-0.021*\\(0.011)} & \makecell{-0.020*\\(0.011)} & \makecell{-0.011***\\(0.002)} & \makecell{-0.014***\\(0.003)} & \makecell{-0.026\\(0.040)} & \makecell{-0.062\\(0.178)} \\
\makecell[l]{\(PD \times\) \\ \(Corruption_{j}\)} & \makecell{-0.011\\(0.012)} & \makecell{-0.010\\(0.013)} & \makecell{-0.010***\\(0.003)} & \makecell{-0.015***\\(0.004)} & \makecell{-0.020\\(0.022)} & \makecell{0.011\\(0.109)} \\
\(PD \times Polity_{i}\) & \makecell{-0.003\\(0.012)} & \makecell{-0.004\\(0.012)} & \makecell{0.010***\\(0.002)} & \makecell{-0.008**\\(0.004)} & \makecell{-0.016\\(0.011)} & \makecell{-0.034\\(0.033)} \\
\(PD \times Polity_{j}\) & \makecell{0.019\\(0.012)} & \makecell{0.020*\\(0.012)} & \makecell{0.008***\\(0.002)} & \makecell{-0.011***\\(0.003)} & \makecell{-0.012\\(0.012)} & \makecell{0.016\\(0.034)} \\
\(N\) & 436{,}706 & 436{,}706 & 436{,}706 & 332{,}911 & 313{,}332 & 155{,}822 \\
Trade data & Trade-Prod & Trade-Prod & Trade-Prod & BACI & Trade-Prod & Trade-Prod \\
\(PD\) data & GDELT & GDELT & GDELT & GDELT & GDELT & GDELT \\
Years & 1980--2020 & 1980--2020 & 1980--2020 & 1995--2023 & 1980--2020 & 1980--2020 \\
\end{longtable}
\endgroup

\begin{posttablenotes}
    (5) and (6) estimated as logit models using the split-panel jackknife method with bootstrapped standard errors based on 1000 trials. All others estimated using PPML with pair-clustered standard errors. All include importer--year, exporter--year and directional pair fixed effects. All except (4) include border--year dummies. * = \(p\) \textless{} 0.1, ** = \(p\) \textless{} 0.05, *** = \(p\) \textless{} 0.01.
\end{posttablenotes}

\vspace{\baselineskip}

\begingroup
\footnotesize
\setlength{\tabcolsep}{4pt}
\renewcommand{\arraystretch}{1.05}

\begin{longtable}{@{}%
  >{\raggedright\arraybackslash}p{0.17\linewidth}%
  >{\centering\arraybackslash}p{0.136\linewidth}%
  >{\centering\arraybackslash}p{0.153\linewidth}%
  >{\centering\arraybackslash}p{0.144\linewidth}%
  >{\centering\arraybackslash}p{0.178\linewidth}%
  >{\centering\arraybackslash}p{0.136\linewidth}@{}}
\caption{Robustness checks on number of TradeProd sectors traded}\label{tbl-rc-sectors}\\
\toprule
& \makecell{(1)\\Drop \\China} & \makecell{(2)\\Drop dem.\\ pairs} & \makecell{(3)\\Drop war\\ pairs} & \makecell{(4)\\Trim \(PD\) \\extremes} & \makecell{(5)\\WGI \\controls} \\
\midrule
\endfirsthead

\toprule
& \makecell{(1)\\Drop China} & \makecell{(2)\\Drop dem. \\pairs} & \makecell{(3)\\Drop war \\pairs} & \makecell{(4)\\Trim \(PD\)\\ extremes} & \makecell{(5)\\WGI \\controls} \\
\midrule
\endhead

\bottomrule
\endlastfoot

\(PD\) & \makecell{-0.088***\\(0.006)} & \makecell{-0.055***\\(0.006)} & \makecell{-0.080***\\(0.008)} & \makecell{-0.098***\\(0.007)} & \makecell{-0.046***\\(0.007)} \\
\(GATTWTO_{1}\) & \makecell{0.259***\\(0.052)} & \makecell{-0.062\\(0.058)} & \makecell{0.232***\\(0.065)} & \makecell{0.234***\\(0.051)} & \makecell{-0.015\\(0.032)} \\
\(GATTWTO_{2}\) & \makecell{0.503***\\(0.100)} & \makecell{-0.004\\(0.113)} & \makecell{0.501***\\(0.125)} & \makecell{0.462***\\(0.099)} & \makecell{0.050\\(0.062)} \\
\makecell[l]{\(PD \times\) \\ \(GATTWTO_{1}\)} & \makecell{0.064***\\(0.006)} & \makecell{0.020***\\(0.006)} & \makecell{0.057***\\(0.008)} & \makecell{0.070***\\(0.007)} & \makecell{0.021***\\(0.007)} \\
\makecell[l]{\(PD \times\) \\ \(GATTWTO_{2}\)} & \makecell{0.075***\\(0.006)} & \makecell{0.031***\\(0.006)} & \makecell{0.071***\\(0.008)} & \makecell{0.083***\\(0.007)} & \makecell{0.037***\\(0.004)} \\
\(N\) & 719{,}224 & 534{,}215 & 479{,}006 & 714{,}984 & 496{,}438 \\
\(PD\) data & UNGA & UNGA & UNGA & UNGA & UNGA \\
Years & 1966--2020 & 1966--2020 & 1966--2020 & 1966--2020 & 1996--2020 \\
\end{longtable}
\endgroup

\begin{posttablenotes}
    All models estimated with pair-clustered standard errors and importer--year, exporter--year, directional pair and border--year fixed effects. All include \(RTA\) and \(PD \times RTA\). Models (1)--(4) also include \(Polity\) and \(Corruption\) (V-Dem) indexes interacted with \(PD\); model (5) replaces these with the WGI Voice and Accountability and Rule of Law. These variables are omitted from the table for brevity. Democratic pairs are those where both countries have \(Polity \geq 6\). (3) drops pairs on either side of, and countries located in, an armed conflict. See Table~\ref{tbl-data} for data sources and details. * = \(p\) \textless{} 0.1, ** = \(p\) \textless{} 0.05, *** = \(p\) \textless{} 0.01.
\end{posttablenotes}

\vspace{\baselineskip}

\begingroup
\footnotesize
\setlength{\tabcolsep}{4pt}
\renewcommand{\arraystretch}{1.05}

\begin{longtable}{@{}%
  >{\raggedright\arraybackslash}p{0.17\linewidth}%
  >{\centering\arraybackslash}p{0.136\linewidth}%
  >{\centering\arraybackslash}p{0.153\linewidth}%
  >{\centering\arraybackslash}p{0.144\linewidth}%
  >{\centering\arraybackslash}p{0.178\linewidth}%
  >{\centering\arraybackslash}p{0.136\linewidth}@{}}
\caption{Robustness checks on number of HS-6 products traded}\label{tbl-rc-products}\\
\toprule
& \makecell{(1)\\Drop \\China} & \makecell{(2)\\Drop dem.\\ pairs} & \makecell{(3)\\Drop war \\pairs} & \makecell{(4)\\Trim \(PD\)\\ extremes} & \makecell{(5)\\WGI\\ controls} \\
\midrule
\endfirsthead

\toprule
& \makecell{(1)\\Drop\\ China} & \makecell{(2)\\Drop dem.\\ pairs} & \makecell{(3)\\Drop war\\ pairs} & \makecell{(4)\\Trim \(PD\)\\ extremes} & \makecell{(5)\\WGI \\controls} \\
\midrule
\endhead

\bottomrule
\endlastfoot

\(PD\) & \makecell{-0.032***\\(0.005)} & \makecell{-0.045***\\(0.006)} & \makecell{-0.021***\\(0.006)} & \makecell{-0.040***\\(0.006)} & \makecell{-0.027***\\(0.004)} \\
\(GATTWTO_{2}\) & \makecell{0.399***\\(0.033)} & \makecell{0.078**\\(0.033)} & \makecell{0.169***\\(0.045)} & \makecell{0.210***\\(0.030)} & \makecell{0.200***\\(0.029)} \\
\makecell[l]{\(PD \times\) \\ \(GATTWTO_{2}\)} & \makecell{0.034***\\(0.005)} & \makecell{0.015***\\(0.006)} & \makecell{0.016**\\(0.007)} & \makecell{0.033***\\(0.006)} & \makecell{0.017***\\(0.004)} \\
\(N\) & 497{,}607 & 337{,}666 & 350{,}020 & 493{,}635 & 791{,}127 \\
\(PD\) data & UNGA & UNGA & UNGA & UNGA & UNGA \\
Years & 1995--2020 & 1995--2020 & 1995--2020 & 1995--2020 & 1996--2023 \\
\end{longtable}
\endgroup

\begin{posttablenotes}
    All models estimated with pair-clustered standard errors and importer--year, exporter--year and directional pair fixed effects. All include \(RTA\) and \(PD \times RTA\). Models (1)--(4) also include \(Polity\) and \(Corruption\) (V-Dem) indexes interacted with \(PD\); model (5) replaces these with the WGI Voice and Accountability and Rule of Law. These variables are omitted from the table for brevity. Democratic pairs are those where both countries have \(Polity \geq 6\). (3) drops pairs on either side of, and countries located in, an armed conflict. See Table~\ref{tbl-data} for data sources and details. * = \(p\) \textless{} 0.1, ** = \(p\) \textless{} 0.05, *** = \(p\) \textless{} 0.01.
\end{posttablenotes}

\vspace{\baselineskip}

\begin{landscape}
\begingroup
\scriptsize 
\setlength{\tabcolsep}{3pt} 
\renewcommand{\arraystretch}{1.1} 

\begin{longtable}{@{}%
  >{\raggedright\arraybackslash\scriptsize}m{0.20\linewidth}%
  *{7}{>{\centering\arraybackslash\scriptsize}m{0.11\linewidth}}@{}}
\caption{Coefficient estimates for number of TradeProd sectors by subperiod}\label{tbl-coefs-sectors}\\
\toprule
& (1) & (2) & (3) & (4) & (5) & (6) & (7) \\
\midrule
\endfirsthead

\toprule
& (1) & (2) & (3) & (4) & (5) & (6) & (7) \\
\midrule
\endhead

\bottomrule
\endlastfoot

\(PD\) & \makecell{-0.017***\\(0.006)} & \makecell{-0.097***\\(0.018)} & \makecell{-0.005\\(0.008)} & \makecell{-0.196***\\(0.022)} & \makecell{-0.059***\\(0.008)} & \makecell{0.074*\\(0.045)} & \makecell{-0.015\\(0.046)} \\
\(RTA\) & \makecell{-0.038***\\(0.013)} & \makecell{0.099***\\(0.015)} & \makecell{0.079***\\(0.017)} & \makecell{-0.053***\\(0.007)} & \makecell{-0.085***\\(0.008)} & \makecell{-0.008***\\(0.003)} & \makecell{0.074***\\(0.023)} \\
\(GATTWTO_{1}\) & \makecell{0.033\\(0.068)} & \makecell{0.061*\\(0.032)} & \makecell{0.004\\(0.039)} & \makecell{0.056\\(0.054)} & \makecell{0.017\\(0.065)} & \makecell{-0.077*\\(0.043)} & \makecell{0.068\\(0.082)} \\
\(GATTWTO_{2}\) & \makecell{0.027\\(0.131)} & \makecell{0.108*\\(0.055)} & \makecell{0.002\\(0.071)} & \makecell{0.139\\(0.105)} & \makecell{0.143\\(0.128)} & \makecell{-0.066\\(0.053)} & \makecell{0.028\\(0.087)} \\
\(PD \times RTA\) & \makecell{0.009*\\(0.006)} & \makecell{0.057***\\(0.012)} & \makecell{0.002\\(0.005)} & \makecell{0.018**\\(0.007)} & \makecell{0.011***\\(0.002)} & \makecell{-0.002\\(0.008)} & \makecell{0.002\\(0.009)} \\
\(PD \times GATTWTO_{1}\) & \makecell{0.009\\(0.006)} & \makecell{0.081***\\(0.019)} & \makecell{0.009\\(0.009)} & \makecell{0.115***\\(0.023)} & \makecell{0.025***\\(0.008)} & \makecell{-0.088*\\(0.047)} & \makecell{-0.010\\(0.048)} \\
\(PD \times GATTWTO_{2}\) & \makecell{0.005\\(0.007)} & \makecell{0.125***\\(0.020)} & \makecell{0.013\\(0.009)} & \makecell{0.166***\\(0.022)} & \makecell{0.058***\\(0.008)} & \makecell{-0.089*\\(0.046)} & \makecell{-0.013\\(0.046)} \\
\(PD \times Corruption_{i}\) & \makecell{-0.003**\\(0.001)} & \makecell{0.006*\\(0.003)} & \makecell{-0.002\\(0.002)} & \makecell{-0.004*\\(0.002)} & \makecell{0.001\\(0.001)} & \makecell{-0.001\\(0.003)} & \makecell{-0.002\\(0.002)} \\
\(PD \times Corruption_{j}\) & \makecell{0.001\\(0.001)} & \makecell{0.009**\\(0.004)} & \makecell{0.001\\(0.001)} & \makecell{-0.002\\(0.003)} & \makecell{-0.003***\\(0.001)} & \makecell{0.000\\(0.003)} & \makecell{0.000\\(0.002)} \\
\(PD \times Polity_{i}\) & \makecell{0.001\\(0.001)} & \makecell{0.003\\(0.002)} & \makecell{0.000\\(0.001)} & \makecell{0.005***\\(0.002)} & \makecell{0.000\\(0.001)} & \makecell{0.003\\(0.003)} & \makecell{-0.003*\\(0.002)} \\
\(PD \times Polity_{j}\) & \makecell{-0.001\\(0.001)} & \makecell{0.007***\\(0.002)} & \makecell{0.000\\(0.001)} & \makecell{0.003\\(0.002)} & \makecell{0.000\\(0.001)} & \makecell{0.000\\(0.003)} & \makecell{0.001\\(0.002)} \\
\(N\) & 109{,}153 & 131{,}133 & 150{,}562 & 168{,}492 & 230{,}835 & 123{,}486 & 179{,}477 \\
\(PD\) data & UNGA & GDELT & UNGA & GDELT & UNGA & GDELT & UNGA \\
Years & 1966--1979 & 1980--1994 & 1980--1994 & 1995--2008 & 1995--2008 & 2009--2020 & 2009--2020 \\
\end{longtable}
\endgroup

\begin{posttablenotes}
    Dependent variable is number of TradeProd sectors. All models include importer--year, exporter--year, directional pair and border--year fixed effects. Pair-clustered standard errors in parentheses. * = \(p\) \textless{} 0.1, ** = \(p\) \textless{} 0.05, *** = \(p\) \textless{} 0.01. 
\end{posttablenotes}

\vspace{\baselineskip}

\begingroup
\footnotesize
\setlength{\tabcolsep}{4pt}
\renewcommand{\arraystretch}{1.05}

\begin{longtable}{@{}%
  >{\raggedright\arraybackslash\scriptsize}m{0.20\linewidth}%
  *{7}{>{\centering\arraybackslash\scriptsize}m{0.11\linewidth}}@{}}
\caption{Coefficient estimates for average trade per sector by subperiod}\label{tbl-coefs-avg}\\
\toprule
& (1) & (2) & (3) & (4) & (5) & (6) & (7) \\
\midrule
\endfirsthead

\toprule
& (1) & (2) & (3) & (4) & (5) & (6) & (7) \\
\midrule
\endhead

\bottomrule
\endlastfoot

\(PD\) & \makecell{0.039*\\(0.021)} & \makecell{-0.142\\(0.087)} & \makecell{0.014\\(0.036)} & \makecell{-0.406***\\(0.097)} & \makecell{-0.080\\(0.049)} & \makecell{0.276\\(0.506)} & \makecell{-0.015\\(0.046)} \\
\(RTA\) & \makecell{0.157***\\(0.030)} & \makecell{-0.046\\(0.067)} & \makecell{-0.038\\(0.068)} & \makecell{0.083**\\(0.040)} & \makecell{0.058\\(0.036)} & \makecell{0.078***\\(0.026)} & \makecell{0.074***\\(0.023)} \\
\(GATTWTO_{1}\) & \makecell{-0.349*\\(0.182)} & \makecell{0.340***\\(0.113)} & \makecell{0.236**\\(0.110)} & \makecell{0.427***\\(0.063)} & \makecell{0.245***\\(0.103)} & \makecell{-0.173\\(0.271)} & \makecell{0.068\\(0.082)} \\
\(GATTWTO_{2}\) & \makecell{-0.285\\(0.194)} & \makecell{0.235*\\(0.127)} & \makecell{0.170\\(0.122)} & \makecell{0.808***\\(0.074)} & \makecell{0.601***\\(0.110)} & \makecell{-0.219\\(0.272)} & \makecell{0.028\\(0.087)} \\
\(PD \times RTA\) & \makecell{0.040***\\(0.009)} & \makecell{-0.112***\\(0.043)} & \makecell{-0.030**\\(0.012)} & \makecell{0.090**\\(0.037)} & \makecell{-0.010\\(0.009)} & \makecell{0.009\\(0.051)} & \makecell{0.002\\(0.009)} \\
\(PD \times GATTWTO_{1}\) & \makecell{-0.036\\(0.025)} & \makecell{0.233**\\(0.106)} & \makecell{0.102*\\(0.058)} & \makecell{0.395***\\(0.106)} & \makecell{0.131***\\(0.049)} & \makecell{-0.473\\(0.505)} & \makecell{-0.010\\(0.048)} \\
\(PD \times GATTWTO_{2}\) & \makecell{-0.057**\\(0.026)} & \makecell{0.198**\\(0.092)} & \makecell{-0.007\\(0.040)} & \makecell{0.475***\\(0.104)} & \makecell{0.076\\(0.050)} & \makecell{-0.522\\(0.508)} & \makecell{0.013\\(0.046)} \\
\(PD \times Corruption_{i}\) & \makecell{0.007**\\(0.004)} & \makecell{-0.002\\(0.011)} & \makecell{-0.004\\(0.005)} & \makecell{0.007\\(0.010)} & \makecell{-0.002\\(0.002)} & \makecell{-0.036**\\(0.016)} & \makecell{-0.002\\(0.002)} \\
\(PD \times Corruption_{j}\) & \makecell{0.002\\(0.003)} & \makecell{-0.007\\(0.011)} & \makecell{0.000\\(0.005)} & \makecell{0.014\\(0.012)} & \makecell{-0.003\\(0.002)} & \makecell{-0.030*\\(0.018)} & \makecell{0.000\\(0.002)} \\
\(PD \times Polity_{i}\) & \makecell{0.005\\(0.003)} & \makecell{-0.004\\(0.013)} & \makecell{0.000\\(0.004)} & \makecell{-0.038***\\(0.013)} & \makecell{-0.002\\(0.003)} & \makecell{0.009\\(0.012)} & \makecell{-0.003\\(0.002)} \\
\(PD \times Polity_{j}\) & \makecell{-0.001\\(0.003)} & \makecell{0.007\\(0.012)} & \makecell{-0.003\\(0.004)} & \makecell{0.015**\\(0.008)} & \makecell{-0.001\\(0.003)} & \makecell{0.010\\(0.020)} & \makecell{0.001\\(0.002)} \\
\(N\) & 109{,}153 & 131{,}133 & 150{,}562 & 168{,}492 & 230{,}835 & 123{,}486 & 179{,}477 \\
\(PD\) data & UNGA & GDELT & UNGA & GDELT & UNGA & GDELT & UNGA \\
Years & 1966--1979 & 1980--1994 & 1980--1994 & 1995--2008 & 1995--2008 & 2009--2020 & 2009--2020 \\
\end{longtable}
\endgroup

\begin{posttablenotes}
    Dependent variable is average trade value in US dollars per TradeProd sector. All models include importer--year, exporter--year, directional pair and border--year fixed effects. Pair-clustered standard errors in parentheses. * = \(p\) \textless{} 0.1, ** = \(p\) \textless{} 0.05, *** = \(p\) \textless{} 0.01.
\end{posttablenotes}

\vspace{\baselineskip}

\newpage

\begingroup
\footnotesize
\setlength{\tabcolsep}{4pt}
\renewcommand{\arraystretch}{1.05}

\begin{longtable}{@{}%
  >{\raggedright\arraybackslash}p{0.12\linewidth}%
  *{6}{>{\centering\arraybackslash}p{0.12\linewidth}}@{}}
\caption{Coefficient estimates on GDELT index by subperiod, political distance only}\label{tbl-gdelt-time}\\
\toprule
& (1) & (2) & (3) & (4) & (5) & (6) \\
\midrule
\endfirsthead

\toprule
& (1) & (2) & (3) & (4) & (5) & (6) \\
\midrule
\endhead

\bottomrule
\endlastfoot

\(PD\) & \makecell{-0.037**\\(0.018)} & \makecell{0.053**\\(0.023)} & \makecell{0.030\\(0.041)} & \makecell{-0.014\\(0.021)} & \makecell{-0.230***\\(0.043)} & \makecell{-0.094***\\(0.034)} \\
\(N\) & 177{,}971 & 157{,}636 & 235{,}934 & 204{,}463 & 269{,}714 & 176{,}537 \\
Trade data & IMF & TradeProd & IMF & TradeProd & IMF & TradeProd \\
\(PD\) data & GDELT & GDELT & GDELT & GDELT & GDELT & GDELT \\
Years & 1980--1994 & 1980--1994 & 1995--2008 & 1995--2008 & 2009--2023 & 2009--2020 \\
\end{longtable}
\endgroup

\begin{posttablenotes}
    Dependent variable is trade in US dollars. All models include importer--year, exporter--year, directional pair and border--year fixed effects. Pair-clustered standard errors in parentheses. * = \(p\) \textless{} 0.1, ** = \(p\) \textless{} 0.05, *** = \(p\) \textless{} 0.01. 
\end{posttablenotes}

\vspace{\baselineskip}

\begingroup
\footnotesize
\setlength{\tabcolsep}{4pt}
\renewcommand{\arraystretch}{1.05}

\begin{longtable}{@{}%
  >{\raggedright\arraybackslash}p{0.10\linewidth}%
  *{8}{>{\centering\arraybackslash}p{0.108\linewidth}}@{}}
\caption{Coefficient estimates on UNGA index by subperiod, political distance only}\label{tbl-unga-time}\\
\toprule
& (1) & (2) & (3) & (4) & (5) & (6) & (7) & (8) \\
\midrule
\endfirsthead

\toprule
& (1) & (2) & (3) & (4) & (5) & (6) & (7) & (8) \\
\midrule
\endhead

\bottomrule
\endlastfoot

\(PD\) & \makecell{-0.037***\\(0.007)} & \makecell{-0.015***\\(0.005)} & \makecell{-0.008\\(0.005)} & \makecell{0.004\\(0.008)} & \makecell{0.001\\(0.004)} & \makecell{-0.002\\(0.003)} & \makecell{-0.011***\\(0.004)} & \makecell{-0.001\\(0.003)} \\
\(N\) & 179{,}890 & 131{,}673 & 218{,}164 & 193{,}937 & 358{,}963 & 292{,}569 & 464{,}326 & 270{,}178 \\
Trade data & IMF & TradeProd & IMF & TradeProd & IMF & TradeProd & IMF & TradeProd \\
\(PD\) data & UNGA & UNGA & UNGA & UNGA & UNGA & UNGA & UNGA & UNGA \\
Years & 1948--1979 & 1966--1979 & 1980--1994 & 1980--1994 & 1995--2008 & 1995--2008 & 2009--2023 & 2009--2020 \\
\end{longtable}
\endgroup

\begin{posttablenotes}
Dependent variable is trade in US dollars. All models include importer--year, exporter--year, directional pair and border--year fixed effects. Pair-clustered standard errors in parentheses. * = \(p\) \textless{} 0.1, ** = \(p\) \textless{} 0.05, *** = \(p\) \textless{} 0.01.
\end{posttablenotes}

\vspace{\baselineskip}

\end{landscape}

\begingroup
\footnotesize
\setlength{\tabcolsep}{4pt}
\renewcommand{\arraystretch}{1.05}

\begin{longtable}{@{}%
  >{\raggedright\arraybackslash}p{0.299\linewidth}%
  *{4}{>{\centering\arraybackslash}p{0.175\linewidth}}@{}}
\caption{Robustness checks with alternative GDELT indexes}\label{tbl-robust-gdelt}\\
\toprule
& \makecell{(1)\\Smaller sample} & \makecell{(2)\\Non-filtered} & \makecell{(3)\\Quarterly} & \makecell{(4)\\First-month} \\
\midrule
\endfirsthead

\toprule
& \makecell{(1)\\Smaller sample} & \makecell{(2)\\Non-filtered} & \makecell{(3)\\Quarterly} & \makecell{(4)\\First-month} \\
\midrule
\endhead

\bottomrule
\endlastfoot

\(PD\) & \makecell{-0.625***\\(0.162)} & \makecell{-0.878***\\(0.137)} & \makecell{-1.329***\\(0.332)} & \makecell{-3.084***\\(0.733)} \\
\(RTA\) & \makecell{0.093**\\(0.038)} & \makecell{0.072**\\(0.031)} & \makecell{0.070**\\(0.034)} & \makecell{0.064*\\(0.033)} \\
\(GATTWTO_{1}\) & \makecell{0.601***\\(0.138)} & \makecell{0.560***\\(0.094)} & \makecell{0.402***\\(0.121)} & \makecell{0.354***\\(0.119)} \\
\(GATTWTO_{2}\) & \makecell{0.914***\\(0.154)} & \makecell{0.860***\\(0.112)} & \makecell{0.711***\\(0.140)} & \makecell{0.674***\\(0.139)} \\
\(PD \times RTA\) & \makecell{0.127**\\(0.059)} & \makecell{0.108*\\(0.064)} & \makecell{0.289**\\(0.134)} & \makecell{0.631**\\(0.305)} \\
\(PD \times GATTWTO_{1}\) & \makecell{0.593***\\(0.155)} & \makecell{0.643***\\(0.134)} & \makecell{1.317***\\(0.331)} & \makecell{3.053***\\(0.745)} \\
\(PD \times GATTWTO_{2}\) & \makecell{0.499***\\(0.161)} & \makecell{0.542***\\(0.138)} & \makecell{1.009***\\(0.335)} & \makecell{2.434***\\(0.738)} \\
\(PD \times Corruption_{i}\) & \makecell{-0.047**\\(0.018)} & \makecell{-0.090***\\(0.022)} & \makecell{-0.107***\\(0.039)} & \makecell{-0.251***\\(0.093)} \\
\(PD \times Corruption_{j}\) & \makecell{-0.020\\(0.020)} & \makecell{-0.043*\\(0.024)} & \makecell{-0.007\\(0.043)} & \makecell{-0.027\\(0.102)} \\
\(PD \times Polity_{i}\) & \makecell{-0.027\\(0.017)} & \makecell{-0.018\\(0.016)} & \makecell{-0.035\\(0.032)} & \makecell{-0.100\\(0.067)} \\
\(PD \times Polity_{j}\) & \makecell{0.004\\(0.016)} & \makecell{0.004\\(0.016)} & \makecell{0.024\\(0.034)} & \makecell{0.050\\(0.071)} \\
\(N\) & 102{,}047 & 732{,}720 & 516{,}049 & 516{,}049 \\
Frequency & Annual & Annual & Quarterly & Quarterly \\
Years & 1980--2020 & 1980--2020 & 1980--2020 & 1980--2020 \\
\end{longtable}
\endgroup

\begin{posttablenotes}
    Dependent variable is trade in US dollars from IMF DOTS. All models include importer--time, exporter--time, directional pair and border--time fixed effects. Pair-clustered standard errors in parentheses. * = \(p\) \textless{} 0.1, ** = \(p\) \textless{} 0.05, *** = \(p\) \textless{} 0.01.
\end{posttablenotes}

\end{document}